\documentclass[%
 reprint,
superscriptaddress,
 amsmath,amssymb,
 aps,
pre,
]{revtex4-1}

\usepackage{graphicx}
\usepackage{epstopdf}
\usepackage{dcolumn}
\usepackage{bm}
\usepackage{esvect}
\usepackage{hyperref}
\usepackage{textcomp}

\newcommand{\unit}[1]{\; \text{#1}}

\begin{document}

\preprint{APS/123-QED}

\title{External control strategies for self-propelled particles: optimizing navigational efficiency in the presence of limited resources}

\author{Daniel F B Haeufle}
\altaffiliation{These authors contributed equally to this work.}
\affiliation{
 Stuttgart Research Center for Simulation Technology, Pfaffenwaldring 5a, Universit\"at Stuttgart, 70569 Stuttgart, Germany
}
\affiliation{
 Institut f\"ur Sport- und Bewegungswissenschaft, Allmandring 28, Universit\"at Stuttgart, 70569 Stuttgart, Germany
}
\author{Tobias B\"auerle}
\altaffiliation{These authors contributed equally to this work.}
\affiliation{
 2. Physikalisches Institut, Universit\"at Stuttgart, Pfaffenwaldring 57, 70569 Stuttgart, Germany
}
\author{Jakob Steiner}
\affiliation{
 2. Physikalisches Institut, Universit\"at Stuttgart, Pfaffenwaldring 57, 70569 Stuttgart, Germany
}
\author{Lena Bremicker}
\affiliation{
 Stuttgart Research Center for Simulation Technology, Pfaffenwaldring 5a, Universit\"at Stuttgart, 70569 Stuttgart, Germany
}
\author{Syn Schmitt}
\affiliation{
 Stuttgart Research Center for Simulation Technology, Pfaffenwaldring 5a, Universit\"at Stuttgart, 70569 Stuttgart, Germany
}
\affiliation{
 Institut f\"ur Sport- und Bewegungswissenschaft, Allmandring 28, Universit\"at Stuttgart, 70569 Stuttgart, Germany
}
\author{Clemens Bechinger}
\affiliation{
 2. Physikalisches Institut, Universit\"at Stuttgart, Pfaffenwaldring 57, 70569 Stuttgart, Germany
}
\affiliation{Max-Planck-Institut f\"ur Intelligente Systeme, Heisenbergstra{\ss}e 3, 70569 Stuttgart, Germany}

\date{\today}

\begin{abstract}
We experimentally and numerically study the dependence of different navigation strategies regarding the effectivity of an active particle to reach a predefined target area. As the only control parameter, we vary the particle's propulsion velocity depending on its position and orientation relative to the target site. By introducing different figures of merit, e.g. the time to target or the total consumed propulsion energy, we are able to quantify and compare the efficiency of different strategies. Our results  suggest, that each strategy to navigate towards a target, has its strengths and weaknesses and none of them outperforms the other in all regards. Accordingly, the choice of an ideal navigation strategy will strongly depend on the specific conditions and the figure of merit which should be optimized.



\end{abstract}

\maketitle

\section{\label{sec:introduction}Introduction}

Self-propelled, i.e., active colloidal particles which are capable to convert energy from their surrounding into directed motion, currently receive considerable attention because they hold use as micron-sized carriers for the delivery of drugs \cite{Drugdelivery,Drugdelivery2} or as active building blocks for the assembly of microsystems \cite{Selfassembly,Selfassembly2}. An important requirement for such applications are efficient steering mechanisms which enable microswimmers to navigate towards specific target sites. In case of e.g. magnetic microswimmers \cite{Magnets1,Magnets2,Magnets3,Magnets4}, whose orientation and propulsion direction is determined by the orientation of an external magnetic field $\mathbf{B}$, navigation is easily achieved by controlling $\mathbf{B}$. In contrast, fully autonomous microswimmers require more complex navigation mechanisms because they are subjected to rotational diffusion. As a consequence, the propulsion direction varies as a function of time. With experiments and simulations, it has been demonstrated that even in the presence of rotational diffusion, autonomous microswimmers can perform a directed motion when moving along planar or patterned walls, funnels, channels or chemical gradients \cite{DiLeonardo,DiLeonardo2,Austin,Simmchen,Hong}. Alternatively, steering of active particles can also be achieved using feedback-loops which control the particle propulsion depending on its position and orientation \cite{Qian2013,Ilic2016}. Employing optical and thermophoretic forces (photon nudging), it has been shown, that high spatial localization of active particles can be achieved \cite{Bregulla2014,Ilic2016}. In contrast to topographical, i.e., static guiding mechanisms, feedback-based control strategies can also be applied in situations with time-dependent target positions.

Here, we investigate with experiments and numerical simulations how the choice of the  control strategy changes the navigation efficiency of self-propelled particles when moving towards a predefined target area. As the only control parameter, we modulate the propulsion velocity, i.e. the particle motility, depending on its current position and orientation by changing the intensity of the homogeneous illumination. Contrary to e.g. phototactic motion of bacteria, it is important to realize, that the particle orientation, being characterized by the rotational diffusion coefficient, is entirely unaffected by our control strategies. We find, that even small changes in the control strategy lead to large differences in the navigation efficiency, the latter being quantified by e.g. the time duration to the target or the total propulsion energy. Because each control strategy has its strengths and weaknesses regarding different figures of merit, the choice of an ideal strategy is not a simple task but strongly depends on the specific navigation problem. In particular, in the presence of external constraints or limited resources, the choice of an optimum control strategy may be crucial for a successful navigation of a particle towards the target site.

\section{\label{sec:methods}Methods}
\subsection{\label{sec:experiment}Experimental setup}
\begin{figure}
	\centering
		\includegraphics[scale=1]{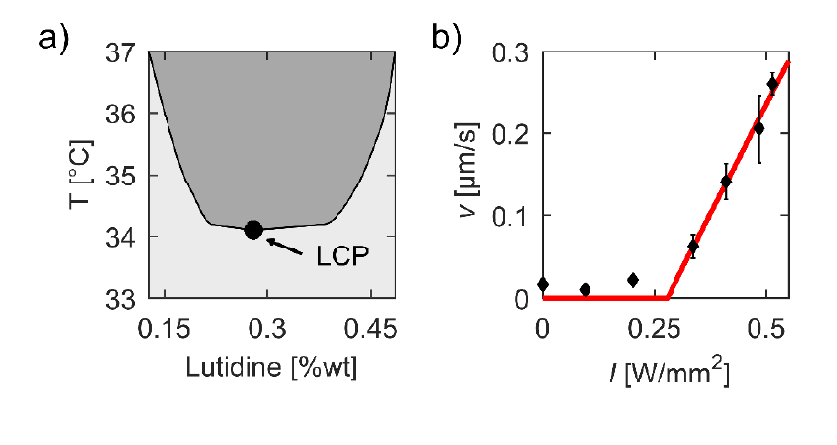}
		\caption{(a) Experimentally determined phase diagram of a water-2,6-lutidine mixture
		with a lower critical point (LCP) at $T_\text{C}=307\unit{K}$. (b) Dependence of the propulsion velocity $v$ on the illumination intensity $I$. The offset corresponds to the threshold intensity below which no demixing of the fluid is induced by laser illumination. Experimental data is shown as symbols and the solid line corresponds to a linear fit above the threshold.}
	\label{fig:setup}
\end{figure}

The microswimmers in this work were fabricated from silica spheres with diameter $\sigma=4.2\unit{\textmu m}$ which were coated on one side with a carbon layer of $30\unit{nm}$ thickness. Such Janus particles were suspended in a binary water-2,6-lutidine mixture that has a lower critical point (LCP) at $T_\text{C}=307\unit{K}$ \cite{waterlutidine} (Fig.\ref{fig:setup}a). Using a thermal bath we set the temperature of our sample cells to $2\unit{K}$ below $T_\text{C}$. 
Illumination of the sample with visible laser light leads to absorption and subsequent heating of the carbon coating. When the temperature of the coating exceeds $T_\text{C}$, the binary liquid becomes locally demixed which leads to propulsion of the particle by self-diffusiophoresis \cite{propulsion1},\cite{propulsion2}. For the illumination intensities $I$ used in this work, the propulsion velocity $v$ linearly increases with $I$ (Fig.\ref{fig:setup}b) above a threshold value $I=0.28\unit{W/mm}^2$. Below this value, the light intensity is not sufficient to induce demixing and no active motion is observed \cite{propulsion1}.

As illumination source, we used a diode-pumped laser with wavelength $\lambda=532\unit{nm}$. To achieve homogeneous (rather than a typical Gaussian profile) illumination over the entire field of view ($390\times310\unit{\textmu m}^2$), the laser beam was scanned over the whole area with an acousto-optical deflector.
At a repetition frequency of $100\unit{kHz}$ the light field can be considered as quasi-static on the relevant timescale. Our experiments were carried out in a thin sample cell where the particles perform a two-dimensional translational and a three-dimensional rotational motion.

\begin{figure}
	\centering
		\includegraphics[scale=1]{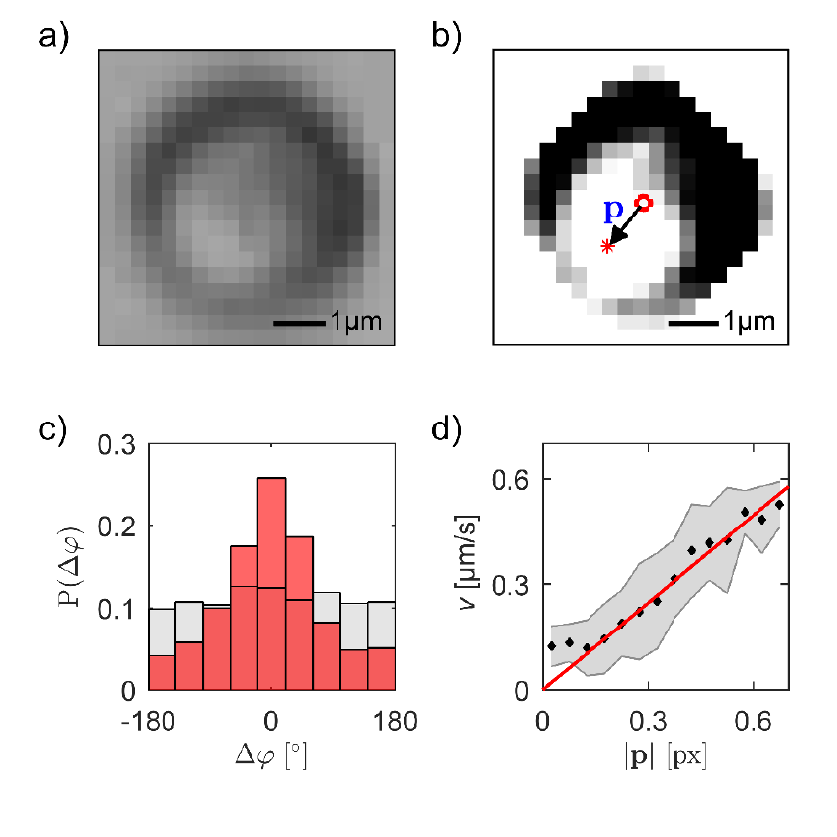}
		\caption{(a) Original image of a Janus particle acquired by a CCD-camera. (b) Image of the particle with increased contrast and calculated barycenter (circle) and intensity centroid (asterisk). The displacement between them points away from the coating and gives the particle orientation $\mathbf{p}$ (arrow). In this image, the position of the intensity centroid is shifted (length of $\mathbf{p}$ increased) to improve visibility.  (c) Probability distribution of the angle $\Delta\varphi$ between $\mathbf{p}$ and the particle velocity $\mathbf{v}$ (light bars: without propulsion, dark (red) bars: with propulsion). (d) Dependence of the particle velocity $v$ on $|\mathbf{p}|$ (linear fit: solid line).}
	\label{fig:cap_orientation}
\end{figure}

For the experimental realization of control strategies it is necessary to obtain and process the time-resolved particle position $\mathbf{r}\left(t\right)$ and its orientation projected to the sample plane $\mathbf{p}\left(t\right)$. To minimize the time delay, we apply an automated analysis protocol, which is optimized for fast processing, immediately after the acquisition of single microscope images (in the following a sampling time of $500\unit{ms}$ is used). First, an intensity threshold filter is applied to the original image (Fig.\ref{fig:cap_orientation}a). This identifies the outer contour of the particle which appears dark under our illumination conditions. After increasing the image contrast, we obtain the barycenter (circle) and the intensity centroid (asterisk) of the particle as schematically shown in Fig.\ref{fig:cap_orientation}b.
In contrast to uncoated particles, where these positions coincide, a characteristic displacement exists for our Janus particles. It is smallest when the cap points up or down and maximal for particle orientations in between. Accordingly, the particle orientation $\mathbf{p}$ (arrow) can be obtained from the vector connecting the barycenter and the intensity centroid.
With this procedure, the entire image analysis takes about $50\unit{ms}$ which can be considered to be instantaneous on the timescale of translational and rotational particle motion (several seconds).

The validity of this approach to determine the particle orientation has been confirmed by comparing $\mathbf{p}$ with the particle propulsion velocity $\mathbf{v}$, which are known to coincide from previous experiments \cite{propulsion3}.
This can be seen in Fig.\ref{fig:cap_orientation}c, where the probability distribution of $\Delta\varphi$ (the angle between $\mathbf{p}$ and $\mathbf{v}$) peaks around $\Delta\varphi=0$ in case of active motion (dark (red) bars) and is almost constant (light bars) without laser illumination, i.e., for Brownian particles, as expected.

Comparing the particle velocity with the length of the particle orientation vector $|\mathbf{p}|$, we find a linear dependence (Fig.\ref{fig:cap_orientation}d, solid line). This confirms that the described image analysis also allows us to reconstruct the three-dimensional particle orientation from projected video images.

Considering the particle position and orientation as input parameters, the particle velocity, i.e., the laser intensity, can be adjusted according to a given control strategy. In the following, we will define three different control strategies which have been realized experimentally.

\subsection{\label{sec:strategies}Control strategies}
Control strategies are employed to navigate a particle from its initial position $\mathbf{r}_0=\left(0,0\right)$ to a target circle with diameter $\sigma_\textbf{T}$ at position $\mathbf{r}_\textbf{T}=\left(l,0\right)$. 
According to them, the desired motility of the particle, i.e. the required illumination intensity, is calculated and controlled as a function of the particle position $\mathbf{r}=\left(x,y\right)$ and orientation $\mathbf{p}$ for each sampling.
The relaxation time to an instantaneous switched on/off laser field has been determined to be less than 10ms. Therefore, the response to a time dependent illumination can be regarded as quasi-instantaneous compared to our typical sampling time of $500\unit{ms}$.

\begin{figure}
\begin{centering}
\includegraphics[width=1\columnwidth]{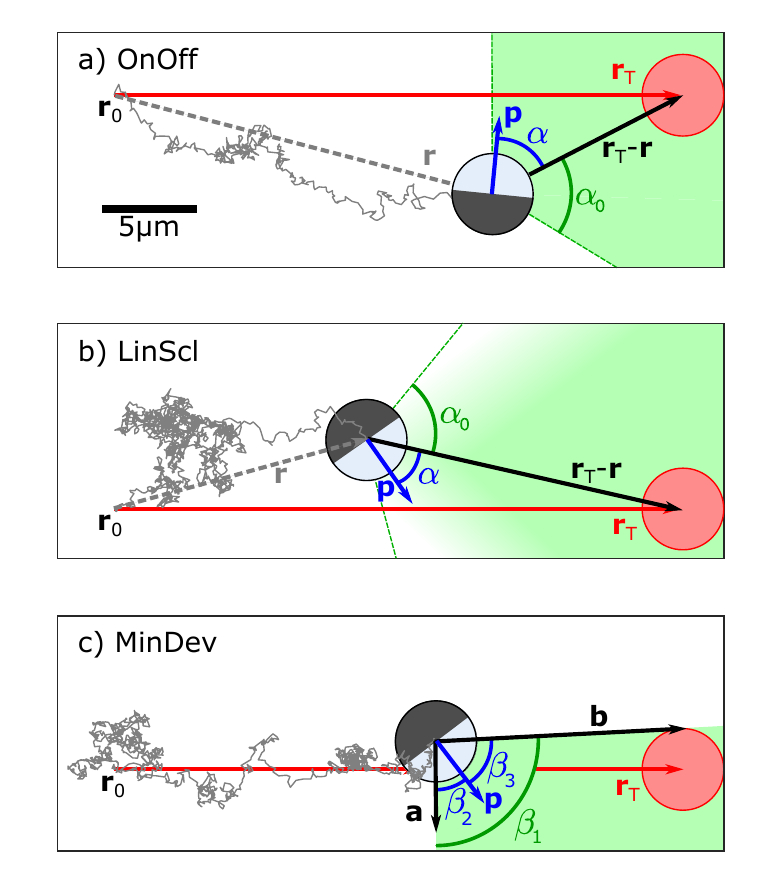}
\end{centering}
		\caption{Schematic description of the different control strategies based on snapshots of simulations (cf. Sec.\ref{sec:simulation}). The particles start at the initial position $\mathbf{r}_0$ and are navigated towards the target area (bullet) around $\mathbf{r}_\text{T}$. The direct connection, i.e., ideal path, is marked by the horizontal line. If the particle orientation $\mathbf{p}$ points towards the shaded area, the propulsion is set to $v=v_\text{max}$ (OnOff (a), MinDev (c)) or to a value between zero and $v_\text{max}$ (LinScl (b)).}
	\label{fig:strategies}
\end{figure}

{\bf OnOff strategy} In this strategy, the propulsion velocity is switched between zero and $v=v_\text{max}$ depending on the angle $\alpha$ between the particle orientation $\mathbf{p}\left(t\right)$ and the vector connecting the current particle position $\mathbf{r}\left(t\right)$ and the target center $\mathbf{r}_\text{T}$ (Fig.\ref{fig:strategies}a):
\begin{equation}
\alpha=\arccos\left(\frac{\mathbf{p}\cdot(\mathbf{r}_\text{T}-\mathbf{r})}{|\mathbf{p}||(\mathbf{r}_\text{T}-\mathbf{r})|}\right).
\end{equation}
When $\alpha$ is smaller or equal than a given value $\alpha_0$, the propulsion velocity is set to $v_\text{max}$ and zero otherwise, i.e.,
\begin{equation}
{v}_{\text{OnOff}}=\begin{cases}
v_\text{max} & \alpha\leq\alpha_{0}\\
0 & \alpha>\alpha_{0}.
\end{cases}
\label{eq:OnOff}
\end{equation}

{\bf {LinScl strategy}} The LinScl (linear scaling) strategy is a variation of the OnOff strategy, where the velocity linearly decreases with $\alpha$ according to
\begin{equation}
{v}_{\text{LinScl}}=\begin{cases}
v_\text{max} \cdot \frac{\alpha_{0}-\alpha}{\alpha_{0}} & \alpha\leq\alpha_{0}\\
0 & \alpha>\alpha_{0}
\end{cases}.
\label{eq:LinScl}
\end{equation}
The velocity is maximal, when the particle points directly towards the target, and zero when the particle orientation deviates by $\alpha_{0}$ and more (Fig.\ref{fig:strategies}b).

{\bf MinDev strategy} The MinDev (minimal deviation) strategy aims at minimizing the particle's distance from the direct straight connection between the initial particle position $\mathbf{r}_{0}$ and the target $\mathbf{r}_\text{T}$ (Fig.\ref{fig:strategies}c).
Given an environment with confined space, e.g. a channel, this external control approach allows reducing the probability of collision and adhesion with the channel walls. In contrast to the previous strategy this requires to store the path or the initial position $\mathbf{r}_0$ in memory for the entire time until the target is reached.

Similar to the OnOff strategy, the propulsion is either set to $v=v_\text{max}$ or $v=0$ depending on whether the particle orientation $\mathbf{p}$ is inside or outside the shaded (green) area in Fig\ref{fig:strategies}c. This area is defined by the two vectors $\mathbf{a}$ and $\mathbf{b}$ where $\mathbf{a}$ is the vertical from the particle center to the connection between $\mathbf{r}_0$ and $\mathbf{r}_\text{T}$ and $\mathbf{b}$ is the shortest connection to the outer contour of the circular target $\mathbf{r}_\text{T}=\left(l,0\right)$ with diameter $\sigma_\text{T} =  \sigma$ (Fig.\ref{fig:strategies}c.),
\begin{eqnarray}
\mathbf{a}&&\ =(0,-\text{sgn}(y)),\\
\mathbf{b}&&\ =\mathbf{r}_\text{T}-\mathbf{r}+(0,\text{sgn}(y)\cdot \sigma_{\text{T}}/2).
\end{eqnarray}
Using
\begin{eqnarray}
\beta_{1}&=\arccos\left(\mathbf{a}\cdot\mathbf{b}/(|\mathbf{a}||\mathbf{b}|)\right),\\
\beta_{2}&=\arccos\left(\mathbf{a}\cdot\mathbf{p}/(|\mathbf{a}||\mathbf{p}|)\right),\\
\beta_{3}&=\arccos\left(\mathbf{b}\cdot\mathbf{p}/(|\mathbf{b}||\mathbf{p}|)\right),
\end{eqnarray}
this leads to the propulsion velocity
\begin{equation}
{v}_{\text{MinDev}}=\begin{cases}
v_\text{max} & \beta_{2}<\beta_{1}\,\,\&\,\,\beta_{3}<\beta_{1}\\
{v}_{\text{OnOff}} & x>\left(l-\sigma_{\text{T}}/2 \right)\,\,\&\,\,x<\left(l+\sigma_{\text{T}}/2\right)\\
0 & \text{else}
\end{cases}
\label{eq:v_MinDev}
\end{equation}
Note, that the middle case in Eq.\ref{eq:v_MinDev} considers the (seldom) events when the particle is directly above or below the target. Then, the area defined by $\mathbf{a}$ and $\mathbf{b}$ becomes very small which makes the control strategy inefficient. 

\subsection{\label{sec:criterie}Figures of merit}
In order to compare the navigation efficiency of different strategies, meaningful quantitative figures of merit are required to measure their performance. The most intuitive figure of merit is probably the time, which may be of essence in future applications either due to a limited lifetime of particles or load they may carry. The MinDev strategy is designed to minimize the deviation from an ideal path, which is therefore a second obvious choice as a figure of merit. Energy may be limited at the source or in the sense of tolerable energy input for the system to avoid damage. At the end, each strategy and each individual control run generates a specific path, each with a different total travel distance for the particle. Based on these considerations, we defined the following four figures of merit.

{\bf Time duration to target } An obvious figure of merit is the total time $T$ required to reach the target area from the initial position $\mathbf{r}_0$. 

{\bf Path deviation } We also calculated the root mean square distance between the particle trajectory and the ideal, i.e., direct path between the initial and the target position
\begin{equation}
D=\sqrt{\frac{1}{N}\sum_{i=1}^{N}y_{i}^{2}},
\end{equation}
where $y_{i}$ corresponds to the vertical particle position at time step $i$ and $N$ is the total number of time steps.

{\bf Propulsion energy } In addition, we computed the total energy which is delivered for the active propulsion. In our case, this is given by the time-accumulated incident laser intensity $I$ 
\begin{equation}
E=\sum_{i=1}^{N}I_{i}\Delta t,
\end{equation}
where  $I_{i}$ is the illumination intensity at time step $i$.

{\bf Total distance }Finally, we determined the total covered distance, i.e., the length of the trajectory 
\begin{equation}
S=\sum_{i=1}^{N-1}\left|\mathbf{r}_{i+1}-\mathbf{r}_{i}\right|.
\end{equation}
Note, that due to the nature of Brownian motion, the measured length of a trajectory will depend on the time steps between the positional measurements.

This is, of course, only a limited selection of possible figures of merit and by no means exhaustive. The chosen quantities relate to resources, such as time, confined navigational space, or energy, which may be limited in future applications of these strategies.

\subsection{\label{sec:simulation}Numerical simulations}

The motion of self-propelled particles is a superposition of stochastic Brownian motion and active propulsion with a given velocity $v$. The Brownian contribution was simulated based on a time discrete evaluation of an independent \textsc{Wiener-Process} for the two translational degrees of freedom $x$ and $y$, and the two rotational degrees of freedom $\varphi$ and $\psi$, representing the rotation around the vertical axis ($\varphi$) and the rotation around an axis, which is perpendicular to the vertical axis and to the symmetry axis of the particle cap ($\psi$). The vertical translation and the rotation around the particle's symmetry axis were ignored, because they will not change the motional behavior and thus play no role in the experiments. The Brownian motion was calculated with pseudo-random variables $\zeta$ with standard normal distribution of zero mean and variance one. The simulated stochastic motion depends on the translational and rotational diffusion constants $D_{T}$ and $D_{R}$, respectively. The differential equation for particle position and orientation was solved for constant time intervals of $\Delta t$
\begin{eqnarray}
\mathbf{r}_{i+1} & = & \mathbf{r}_{i}+\mathbf{\zeta_{r}}\sqrt{2D_{T}\Delta t}+\Delta\mathbf{r}_{\text{active}}(\mathbf{r}_{i},\varphi_{i},\psi_{i}),\\
\varphi_{i+1} & = & \varphi_{i}+\zeta_{\varphi}\sqrt{2D_{R}\Delta t},\\
\psi_{i+1} & = & \psi_{i}+\zeta_{\psi}\sqrt{2D_{R}\Delta t},
\label{eq:dgl}
\end{eqnarray}
with the active particle movement
\begin{equation}
\Delta\mathbf{r}_{\text{active}}=\left(\begin{array}{c}
\cos(\varphi_{i})\\
\sin(\varphi_{i})
\end{array}\right)\cos(\psi_{i})\sqrt{2}v(\mathbf{r}_{i},\mathbf{p}_{i})\Delta t.
\label{eq:ractive}
\end{equation}
The mean horizontal velocity $v$ which is set according to the navigation strategies (Eq., \ref{eq:OnOff}, \ref{eq:LinScl}, \ref{eq:v_MinDev}), depends on the current particle position $\mathbf{r}_i$ and the  in-plane particle orientation $\mathbf{p}_i$, which is given by
\begin{equation}
\mathbf{p}_i=\left(\begin{array}{c}
\cos(\varphi_{i})\\
\sin(\varphi_{i})
\end{array}\right)\cos(\psi_i),
\end{equation}
with  $|\mathbf{p}|=\cos(\psi_i)$.
The factor $\sqrt{2}$ in Eq.\ref{eq:ractive} takes into account, that the particle can rotate freely in all three dimensions.

To obtain sufficient statistics, we simulated 600 runs for each control strategy and determined the mean and the standard deviation of the four figures of merit. 
From our simulations and using statistical power analysis software \footnote{Software: G*Power, Heinrich-Heine University Duesseldorf}, we estimated the number of experimental data points to obtain representative results.

\subsection{\label{sec:parameteridentification} Simulation Parameters}
To use realistic parameters in our simulations, we used the experimentally determined values for the translational $D_{T}$ and rotational $D_{R}$ diffusion coefficients. For our colloidal particles ($\sigma=4.2\unit{\textmu m}$), we obtained $D_{T}=0.027\unit{\textmu m}^2\text{/s}$ and $D_{R}=1/120 \unit{s}^{-1}$. While $D_{R}$ is in almost perfect agreement with the corresponding Stokes-Einstein value, $D_{T}$ is about 50\% below the theoretical value. Such behavior is due to hydrodynamic interactions \cite{Brenner} with the wall and in good agreement with previous studies \cite{propulsion3}.

\begin{figure}
	\centering
		\includegraphics[width=1\columnwidth]{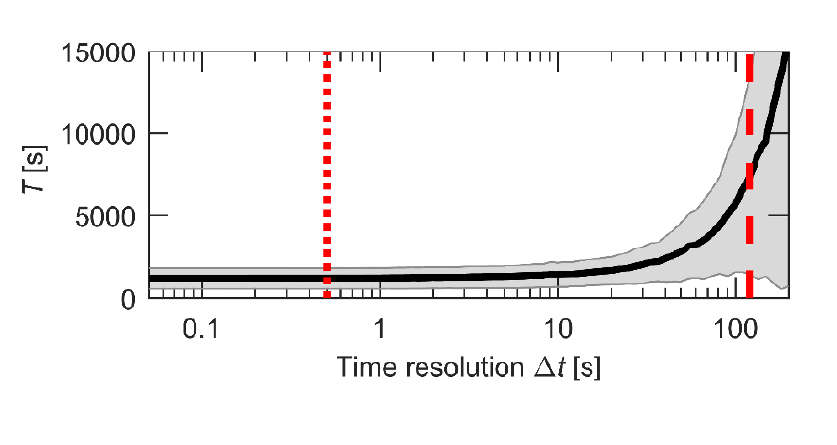}
		\caption{Simulated time duration to target $T$ with OnOff $63^\circ$ strategy as a function of the time resolution $\Delta t$. With increasing $\Delta t$, $T$ strongly increases. In our experiments and simulations, we have chosen $\Delta t=0.5\unit{s}$ (vertical dotted line). In this regime, no influence of $\Delta t$ on $T$ and the other figures of merit is observed. The vertical dashed line corresponds to the particle's rotational diffusion time.}
	\label{fig:TimeResolution}
\end{figure} 
 
The temporal resolution $\Delta t$ of the positional and orientational data acquisition and the feedback loop is an important parameter because it strongly affects how fast the control parameter responds to a change in the particle configuration. Accordingly, the above defined figures of merit will strongly depend on $\Delta t$. This is exemplarily shown in Fig.\ref{fig:TimeResolution} where we plot the time duration to target $T$ for the OnOff $63^\circ$ strategy and a target distance $l=30\unit{\textmu m}$ as a function of $\Delta t$. As expected $T$ increases for large $\Delta t$ and saturates towards small $\Delta t$. Below $\Delta t\approx 3\unit{s}$, $T$ becomes constant. For other strategies and figures of merit a similar behavior is observed. In the following, we set $\Delta t = 0.5 \unit{s}$ in the simulations and the experiments.

\begin{figure}
	\centering
		\includegraphics[width=1\columnwidth]{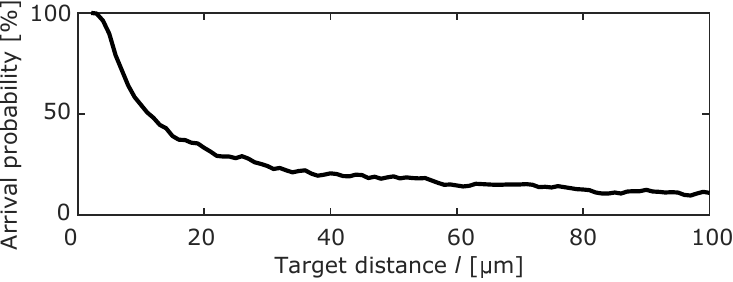}
		\caption{
		Arrival probability of uncontrolled particles (simulation) within six hours for different target distances $l$. With pure Brownian diffusion (dotted line), the probability is only about $15\%$ for $l=30\unit{\textmu m}$ and below $1/600$ for $l=100\unit{\textmu m}$. The probability rises for active particles with permanent homogeneous illumination (solid line) to about $25\%$ for $l=30\unit{\textmu m}$ and approx. $10\%$ for $l=100\unit{\textmu m}$. With active control (OnOff strategy, dashed line), all particles reach the target area. }
	\label{fig:AlwaysOnVarDistance}
\end{figure}

So far, we have not yet specified the distance $l$ between the initial particle position and the target. The choice of this distance strongly affects the typical duration of a single run in the experiments and the simulations. When $l$ is too large, the time to reach the target will increase. On the other hand, when $l$ is too small, a considerable number of active particles will reach the target even in absence of a control strategy. This is demonstrated in Fig.\ref{fig:AlwaysOnVarDistance}, where we show the probability that an active particle reaches the target at distance $l$ without a control strategy, i.e., under constant illumination conditions, within six hours. 
In the following, we have set $l=30\unit{\textmu m}$ and $l=100\unit{\textmu m}$ in our experiments and simulations because the chance to reach the target without control is below $25\%$ and approx. $10\%$, respectively.

\section{Results}
For a target distance $l=30\unit{\textmu m}$ we investigated five different control strategies: OnOff with $\alpha_0=\{13^\circ, 32^\circ, 63^\circ\}$, LinScl with $\alpha_0=63^\circ$ and MinDev.

\begin{figure}
	\centering
	\includegraphics[width=1\columnwidth]{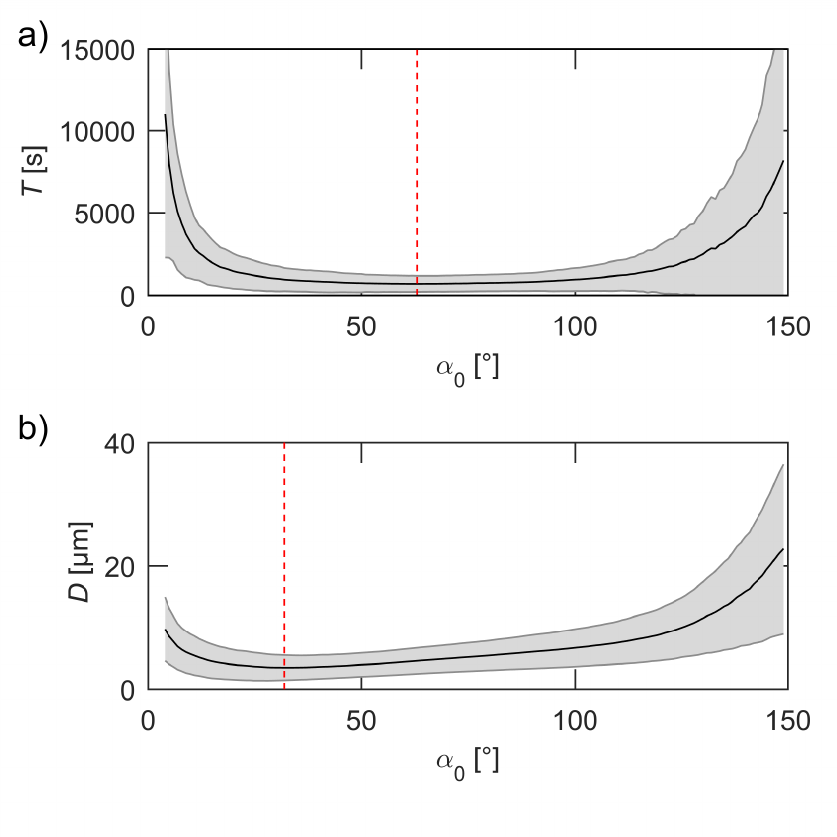}
	\caption{
			{Performance of the OnOff strategy depending on the angle $\alpha_0$ (see eq. \ref{eq:OnOff}) for two of the figures of merit: (a) time duration to target $T$, (b) path deviation $D$.
		Shown are the simulation results for a target distance of $l=30\unit{\textmu m}$.
		The global minima (dashed lines) are at $\alpha_{0}=63 ^\circ$ for $T$ and $\alpha_{0}=32 ^\circ$ for $D$.}
	}
	\label{fig:alpha0dep}
\end{figure}

The choice of the $\alpha_0$-values was a result of our simulations which showed that OnOff $63^\circ$ minimizes the time duration $T$ and OnOff $32^\circ$ minimizes the path deviation $D$. Because $T$ strongly increases with decreasing  $\alpha_0$, the choice $\alpha_0=13^\circ$ corresponds to the smallest value which is still accessible on experimental time scales.
In all experiments and simulations shown below, the maximum particle velocity was set to $v_\text{max}=0.2\unit{\textmu m/s}$.
To allow for a direct comparison between experiments and simulations, the initial particle orientation $\mathbf{p}$ was always chosen to be pointing towards $\mathbf{r}_\text{T}$.

\begin{figure}
	\centering
		\includegraphics[scale=1]{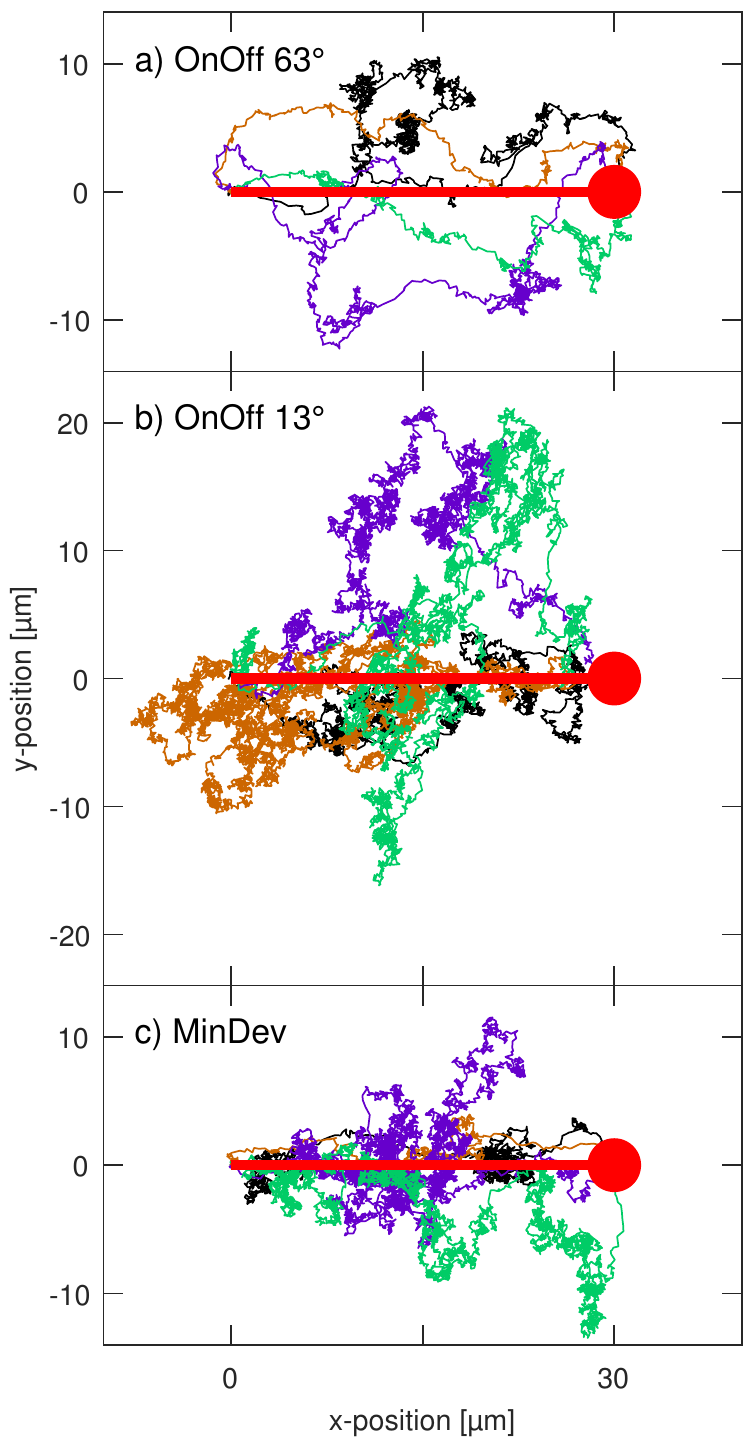}
		\caption{Experimental trajectories for (a) OnOff $63^\circ$, (b) OnOff $13^\circ$ and (c) MinDev strategy. The ideal path from start to target is marked as a solid straight line with the circular target area in real proportion.}
	\label{fig:trajectories30}
\end{figure}

\begin{figure}
	\centering
	\includegraphics[scale=1]{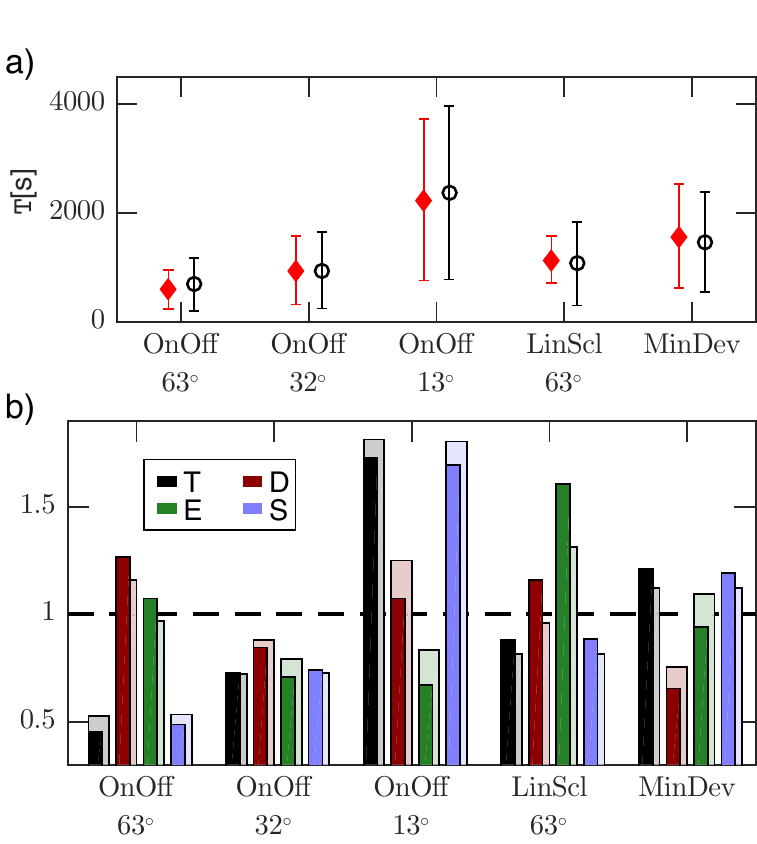}
	\caption{(a) Experimental (closed symbols) and numerically (open symbols) determined time duration to target $T$ for different control strategies. (b) Summary of the performance of different control strategies for a target distance of $l=30\unit{\textmu m}$ (experimental and numerical data as filled and open bars, respectively. 
	 The figures of merit are normalized by the mean value of the respective criterion.}
	\label{fig:fingerprint30}
\end{figure}

At first, we want to mention, that for all control strategies discussed here, in the experiments all particles reached the target within less than six hours (cf. Fig.\ref{fig:fingerprint30}a). Fig.\ref{fig:trajectories30} exemplarily shows some experimental trajectories for three of the above discussed control strategies. The direct path and the finite sized target area are marked as solid line and a bullet, respectively. Note, that the latter region is shown in real scale.
Comparing the $\alpha_0$-dependence of the OnOff strategies (Fig.\ref{fig:trajectories30}a,b), one recognizes that with decreasing $\alpha_0$ the trajectory exhibits an increasing fraction of segments with mere Brownian motion. This is due to the decreasing probability that the particle's orientation meets the condition where the propulsion is switched on. In case of the OnOff $13^\circ$ strategy (Fig.\ref{fig:trajectories30}b), the Brownian motion dominates which results in large particle excursions from the direct path. This is in strong contrast to the MinDev strategy, where the trajectories are much more confined close to the direct path to the target although they also contain a large number of Brownian parts. Compared to the OnOff-strategy, where the particle can propel even away from the direct path, this is hardly possible in the MinDev strategy (cf. Fig.\ref{fig:strategies}).

For further analysis we take a closer look at the figures of merit defined above. Fig.\ref{fig:fingerprint30}a shows, that the time $T$ to reach the target varies by more than a factor of three between the investigated control strategies. The shortest time is achieved with OnOff $63^\circ$ and increases with decreasing $\alpha_0$. The LinScl $63^\circ$ and MinDev perform worse than OnOff $63^\circ$ and OnOff $32^\circ$. We find excellent agreement between experimental (open symbols) and simulation data (closed symbols), not only regarding the mean but also with respect to the standard deviation of $T$. The results for the total distance $S$ (data not shown) show a very similar dependence on the different control strategies. In contrast, the path deviation $D$ and the propulsion energy $E$ vary by less than a factor of two between the different strategies (data not shown).

Fig.\ref{fig:fingerprint30}b compares all figures of merit for the different strategies. Experimental and numerical data are shown as opaque and transparent bars, respectively. For a better comparison, we have normalized the data such, that the mean value of each criterion, averaged over all strategies, is one. As a consequence, any value below (above) one indicates good (bad) relative performance regarding a chosen criterion.
For example, OnOff $63^\circ$ is best regarding $T$ and $S$, but worse when considering $D$ and $E$. In contrast, OnOff $32^\circ$ performs equally well at all figures of merit. MinDev optimizes the path deviation $D$ at the expense of the performance regarding all other figures of merit. 
This synopsis allows to recognize the benefits and drawbacks of the strategies. It highlights the fact, that there is no ``best'' strategy because each of them has its advantages and disadvantages.

\begin{figure}
	\centering
	\includegraphics[scale=1]{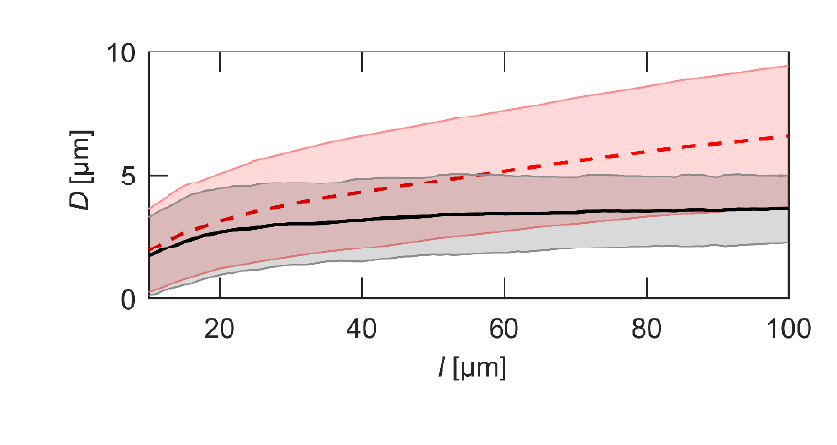}
	\caption{Simulations demonstrating the influence of $l$ on $D$ for the control with MinDev (solid) and OnOff $\alpha_0$ (dashed). The shaded areas indicate the standard deviation. In the OnOff strategy $\alpha_0$  depends on $l$ and is chosen to minimize $D$ at each distance $l$. For small values of $l$ both strategies do not differ much in $D$, whereas for large $l$ the advantage of MinDev starts to gain importance, resulting in a smaller increase than for OnOff.}
	\label{fig:Doverl}
\end{figure}

The figures of merit will also depend on the value of the target distance $l$. Simulations predict, that the differences between the strategies grow for increasing target distance. This is exemplarily shown for the path deviation $D$ (Fig.\ref{fig:Doverl}). For small distances, the mean values of $D$ are similar for the OnOff and MinDev strategies. For increasing $l$, $D$ almost saturates for MinDev but continues to grow significantly in the case of the OnOff strategy.

\begin{figure}
	\centering
		\includegraphics[scale=1]{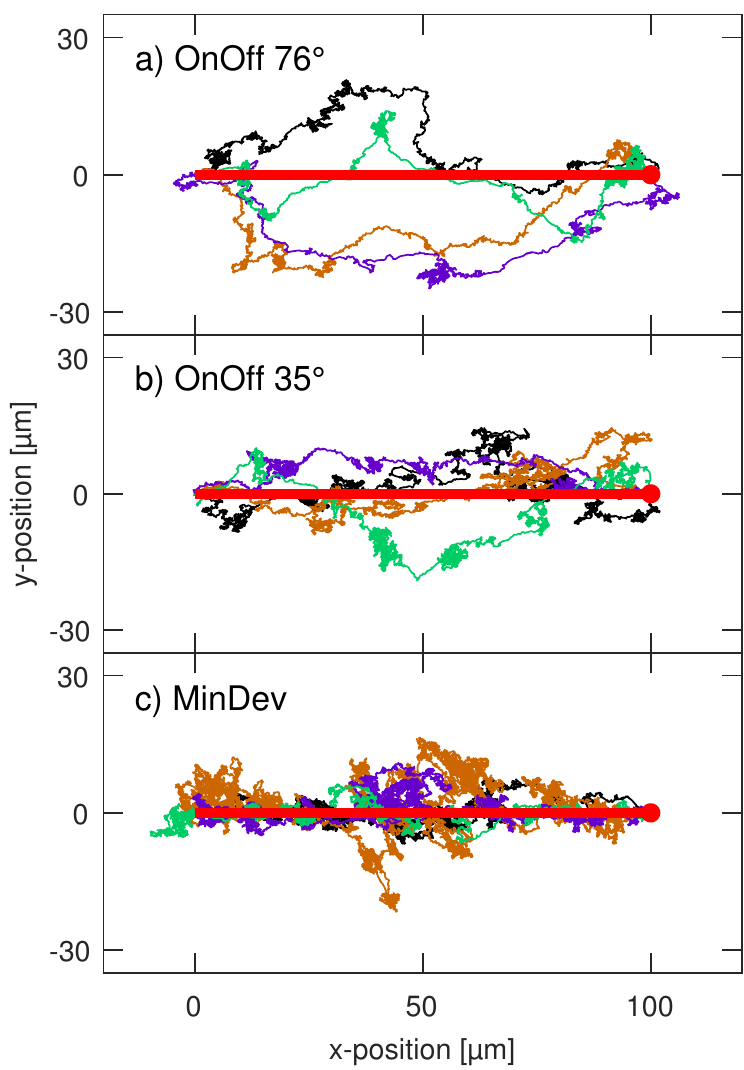}
		\caption{Experimental trajectories for (a) OnOff $76^\circ$, (b) OnOff $35^\circ$ and (c) MinDev and a target distance $l=100\unit{\textmu m}$. }
	\label{fig:trajectories100}
\end{figure}

In the following, we performed experiments with $l=100\unit{\textmu m}$ to test these predictions and to confirm the expected advantage of the MinDev strategy.
Three strategies were investigated: OnOff $76^\circ$ (minimizing $T$), OnOff $35^\circ$ (minimizing $D$) and MinDev (designed to minimize $D$). The trajectories shown in Fig.\ref{fig:trajectories100} highlight, that in case of MinDev, the particle remains close to the ideal path and never actively propels away. For the OnOff strategies, however, the particles can actively propel further away from the ideal path.

\begin{figure}
	\centering
		\includegraphics[scale=1]{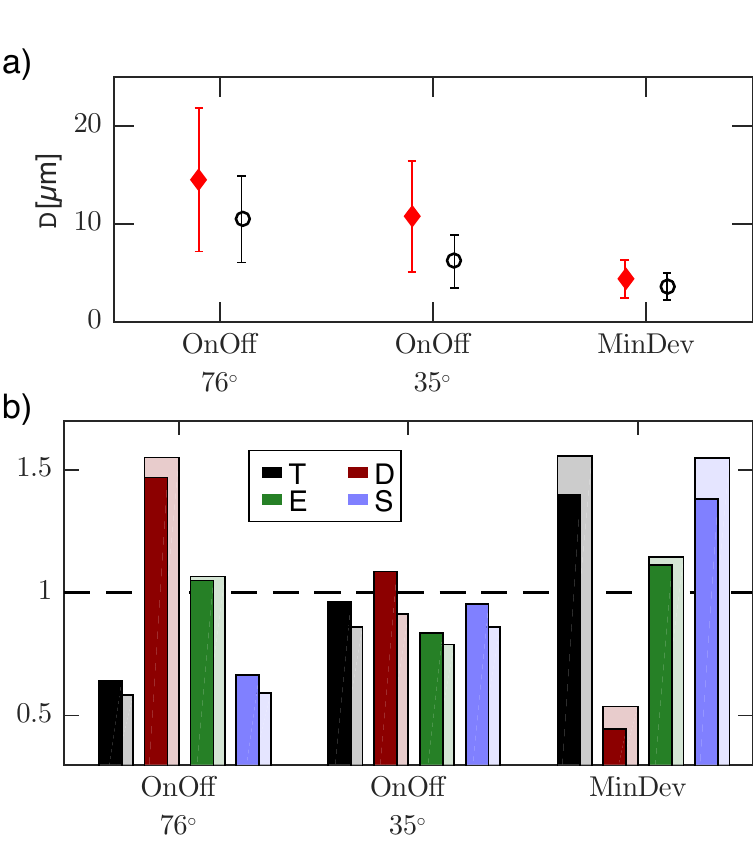}
		\caption{(a) Experimental (closed symbols) and numerically (open symbols) determined path deviation $D$ for different control strategies. (b) Summary of the performance of different control strategies for a target distance of $l=100\unit{\textmu m}$ (experimental and numerical data as filled and open bars, respectively. 
			The figures of merit are normalized by the mean value of the respective criterion.}☺
	\label{fig:fingerprint100}
\end{figure}

As expected, the path deviation $D$ is significantly smaller for the MinDev strategy than for the other strategies (Fig.\ref{fig:fingerprint100}a). In addition, the standard deviation becomes quite small. Obviously, for larger target distances, the MinDev strategy provides an efficient steering mechanism where the particles hardly deviate from the ideal, i.e., straight, path to the target. Also, the differences regarding the figures of merit in the different strategies become more pronounced for $l=100\unit{\textmu m}$ (Fig.\ref{fig:fingerprint100}b). OnOff $76^\circ$ performs very well regarding $T$ and $S$, but badly regarding $D$. OnOff $35^\circ$ shows a good trade-off between all figures of merit. MinDev optimizes $D$ at the expense of all other figures of merit.

\subsection*{\label{sec:RotationRate}Control of the rotation rate}

\begin{figure}
	\centering
		\includegraphics[width=.85\columnwidth]{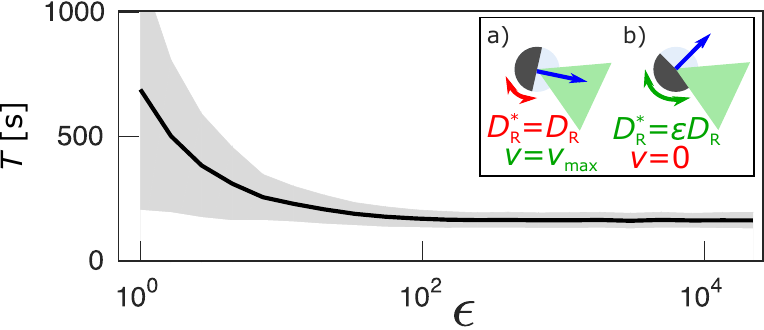}
		\caption{Influence of the diffusional enhancement factor $\epsilon$ for OnOff $63^\circ$ and $l=30\unit{\textmu m}$ on the time duration to target $T$. Increasing the random reorientation rate improves $T$ by more than a factor two. 
		The inset schematically illustrates the strategy: (a) when the particle is oriented towards the target, the particle has exhibits the Stokes-Einstein rotational diffusion $D_{R}$ and is propelled with $v=v_{\max}$; (b) when the particle is not oriented towards the target, the rotational diffusion is increased by a factor $\epsilon$ and the propulsion is deactivated ($v=0$).}
	\label{Fig:varActiveRot}
\end{figure}

So far, we have only varied the propulsion velocity to steer the particles towards a target region. From a theoretical point of view, the navigation efficiency can be enhanced when we additionally introduce a variable rate at which the particle can reorient itself. For colloidal swimmers the reorientation rate is given by their rotational diffusion constant $D_\text{R}$ which is fully determined by the particle size and the solvent’s viscosity. Accordingly, such variations of $D_\text{R}$ have currently not been realized. However, it has been theoretically predicted that the orientation of active particles can be adjusted by light gradients \cite{Wurger}. Furthermore, this concept is interesting from a biological point of view, as many motile organisms
(e.g. E. coli \cite{BergBraun,Berg2000}) can actively change their orientation by reversing the flagella motion which eventually leads to a "run and tumble" motion. 
Inspired by this biological strategy, we have modified our control strategies such, that the rotation coefficient of a particle will depend on its current orientation and, thus, on its propulsion velocity. To enhance the navigation efficiency, we assume, that the effective rotational diffusion coefficient $D_{R}^*$ becomes larger when the particle is not self-propelling and smaller otherwise.
\begin{equation}
D^{*}_{R} =\begin{cases}
\epsilon D_R & v=0\\
D_R & v>0.
\end{cases}
\end{equation}
Accordingly, when the particle is oriented in a {\emph{wrong}} direction where self-propulsion is turned off ($v=0$), it will rotate faster ($\epsilon>1$). This increases the chance that the particle reorients towards a favorable direction where self-propulsion is turned on (see inset Fig.\ref{Fig:varActiveRot}). Then, $D_{R}^*$ becomes smaller again. To implement this strategy in our simulations, $D_R$ is replaced by $D_{R}^*$ in Eq.(\ref{eq:dgl}).

Fig.\ref{Fig:varActiveRot} shows the dependence of the time duration to the target $T$ as a function of the diffusional enhancement factor $\epsilon$. Clearly, the additional variation of the rotational diffusion coefficient leads to a substantial improvement in $T$ and the other figures of merit (not shown). For $\epsilon>150$, however, the performance saturates. At these values the rotation rate becomes so high, that the particle can fully rotate within a single time step. 

\section{\label{sec:Summary}Summary}
In our study, we have demonstrated how the choice of a specific control strategy affects the navigation of self-propelled particles towards a predefined target site. To quantify the navigation efficiency, we have computed different figures of merit e.g. the time duration to target or the total energy required to reach the target. Our results show, that 
each of the control strategies has its strengths and weaknesses, and none of them outperforms the others in all regards. In addition, we have demonstrated, that the optimal strategy also depends on the target distance. 
Because the motional behavior of active particles is rather independent of the specific driving mechanism of micron-sized objects, we expect that the navigation strategies discussed here will also apply to other systems which eventually may find use as e.g. drug delivery systems where time, accessible space, provided energy, etc. are limited.

\section*{Acknowledgments}
We thank Celia Lozano, Felix K\"ummel, and Borge ten Hagen for fruitful discussions and input to our study. We want to acknowledge financial support from the Deutsche Forschungsgemeinschaft within the Priority Programme Microswimmers SPP 1726.

\bibliography{Papers}

\end{document}